\documentclass[aps,pra,twocolumn,superscriptaddress,longbibliography]{revtex4}
\usepackage[usenames,dvipsnames]{color}
\usepackage{tikz}
\usepackage{graphicx}
\usepackage{dcolumn}
\usepackage{bm}
\usepackage{amsmath}
\usepackage{amsfonts}
\usepackage{psfrag}
\usepackage[bookmarks,colorlinks=false]{hyperref}
\usepackage{mathtools}
\usepackage{accents}

\usepackage[mathcal]{euscript}

\begin{document}

\title{Hybrid Longitudinal-Transverse Phonon Polaritons}

\author{Christopher R. Gubbin}
\affiliation{School of Physics and Astronomy, University of Southampton, Southampton, SO17 1BJ, United Kingdom}
\author{Rodrigo Berte}
\affiliation{CAPES Foundation, Ministry of Education of Brazil, Brasilia, DF 70040-020, Brazil}
\author{Michael A. Meeker}
\author{Alexander J. Giles}
\author{Chase T. Ellis}
\author{Joseph G. Tischler}
\author{Virginia D. Wheeler}
\affiliation{U.S. Naval Research Laboratory, Washington, DC 20375 United States}
\author{Joshua D. Caldwell}
\affiliation{Department of Mechanical Engineering, Vanderbilt University, Nashville, Tennessee 37205 United States}
\author{Simone De Liberato}
\email[Corresponding author: ]{s.de-liberato@soton.ac.uk}
\affiliation{School of Physics and Astronomy, University of Southampton, Southampton, SO17 1BJ, United Kingdom}

\begin{abstract}
We demonstrate how to exploit long-cell polytypes of silicon carbide to achieve strong coupling between transverse phonon polaritons and zone folded longitudinal optical phonons. The resulting quasiparticles possess hybrid longitudinal and transverse nature, allowing them to be generated through electric currents while emitting radiation to the far field. 
We develop a microscopic theory predicting the existence of the hybrid longitudinal-transverse excitations. We then provide their first experimental observation by tuning the monopolar resonance of a nanopillar array through the folded longitudinal optical mode, obtaining a clear spectral anti-crossing. This represents an important first step in the development of electrically pumped mid-infrared emitters.
\end{abstract}

\maketitle

\section*{Introduction}
Phonon polaritons are mixed light-matter excitations arising from hybridisation between photons and transverse optical phonons in polar dielectrics. In the crystal's Reststrahlen band, between the transverse optical (TO) and longitudinal optical (LO) phonon frequencies, the real part of the dielectric function is negative. In this spectral window these resonances can be strongly localised at the crystal surface leading to the appearance of localised modes termed surface phonon polaritons (SPhPs).\\ 
Following initial studies on surfaces \cite{Hillenbrand2002,Greffet2002} or in waveguides \cite{Holmstrom2012}, localization of SPhPs  in user defined nanostructures was achieved \cite{Caldwell2013,Wang2013}. 
Such localised excitations allow for an energy confinement on length-scales orders of magnitude shorter than that of the free photon wavelength \cite{Caldwell2015}, without the strong optical losses associated with plasmonic systems \cite{Khurgin2015, Khurgin2018}. 
They are also extremely tunable, thanks to their morphologic nature \cite{Gubbin2017}, their dependence on carrier density \cite{Dunkelberger2018,Spann2016} and their ability to hybridise with propagating \cite{Gubbin2016} or epsilon-near-zero modes \cite{Passler2018}. 
Recent investigations have demonstrated the potential of localised phonon polaritons for sensing \cite{Berte2018}, nonlinear optics \cite{Gubbin2017b, Gubbin2017c, Razdolski2016a, Razdolski2018}, waveguiding \cite{Alfaro-Mozaz2017}, nanophotonic circuitry \cite{Li2018}, and rewritable nano-optics 
\cite{Li2016,Folland2018}.\\
The tunable, narrowband nature of SPhP resonances makes them a good candidate for realisation of integrated mid-infrared emitters. To this end phonon polariton-based thermal emitters have been demonstrated \cite{Wang2017, Schuller2009,Greffet2002}, but thermal pumping is intrinsically inefficient and does not allow for an increase in the degree of temporal coherence.
In polaritonic systems based on electronic excitations, electrical injection of polaritonic modes has been demonstrated \cite{Schneider2013}, but similar schemes with phonon polaritons are difficult to implement as their energies typically lie an order of magnitude below the electrical bandgap. Electrical currents do however couple efficiently to crystal lattice vibrations. In fact one of the main sources of Ohmic loss in polar dielectrics is through LO phonon emission via the Fr{\"o}hlich interaction. 
The use of such an interaction to pump SPhPs is however problematic as the LO phonon frequency defines the upper edge of the Reststrahlen band, which limits the spectral overlap and resonant transitions between the LO phonon and SPhPs. More fundamentally the photonic field, due to its transverse nature, doesn't couple with longitudinal excitations. We are thus at an impasse: we can efficiently create large populations of LO phonons via electrical currents, but only the TO ones can form SPhP and emit light in the far-field. 
\begin{figure}
	\centering
	\includegraphics[width=\columnwidth]{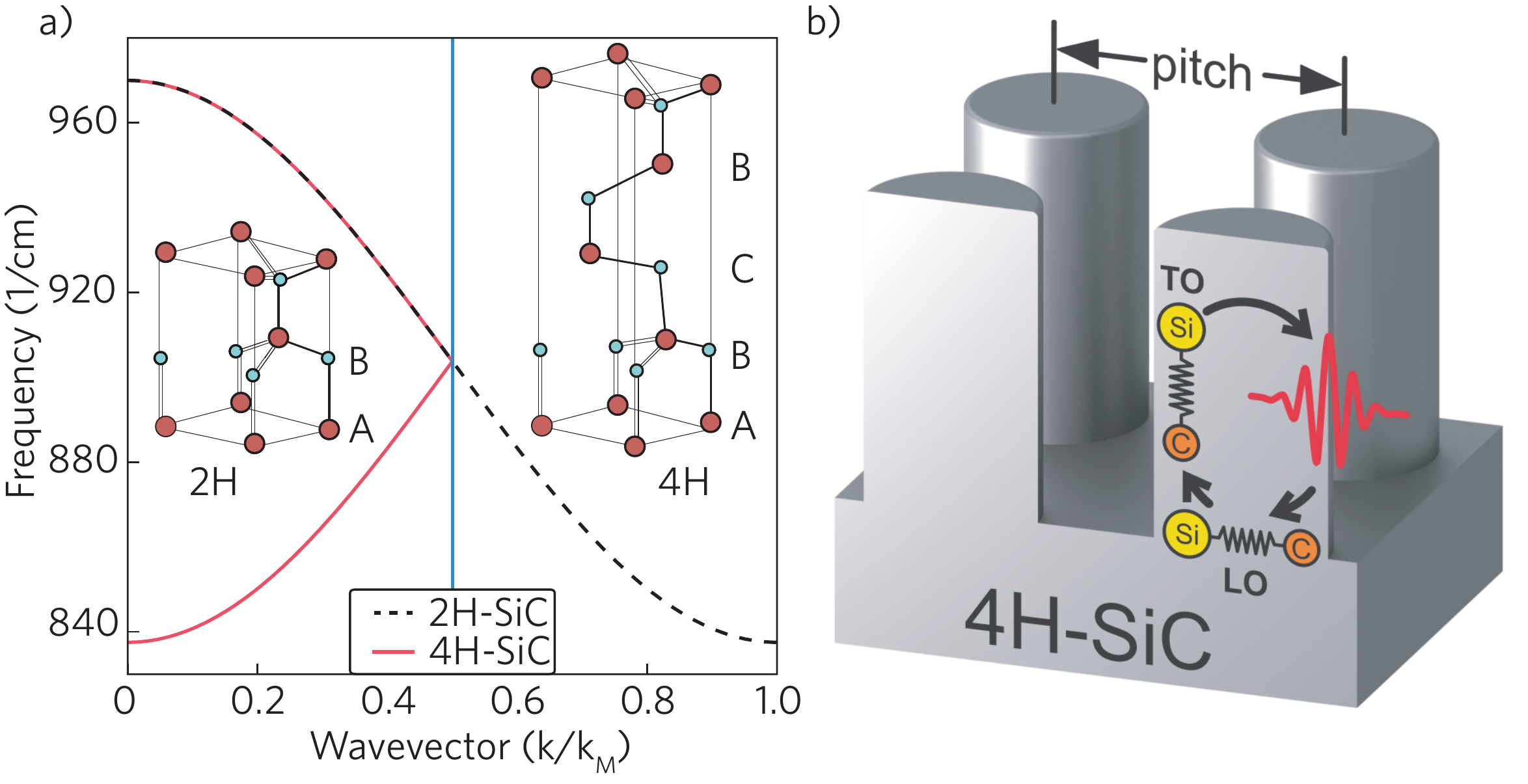}
	\caption{\label{fig:Fig1} a) Illustration of LO phonon dispersion parallel to the c-axis in 2H- and 4H-SiC \cite{Nakashima1997}. The wavevector is normalised  over the 2H-SiC Brillouin zone border $k_{\mathrm{M}}$. Inset shows an illustration of the crystal structures of 2H- and 4H-SiC, the length of the 4H- unit cell is approximately twice that of the 2H. b) Sketch of the strong coupling between LO phonons, TO phonons, and photons, resulting in the creation of LTPPs, in a square array of 4H-SiC nanopillars.}
\end{figure}\\
In this paper we demonstrate how these problems can be solved,  exploiting Bragg zone-folded LO phonons (ZFLO) in silicon carbide (SiC) polytypes whose unit cells are elongated along the c-axis.  This allows us to engineer and observe novel hybrid modes arising from the strong coupling of transverse SPhP modes and ZFLO phonons. The resulting quasiparticles, which we name Longitudinal-Transverse Phonon Polaritons (LTPP) possess both a longitudinal character, allowing resonant generation by Ohmic losses, and a transverse one making far-field emission possible. The ZFLO modes we exploit, typically termed weak phonons, are high-wavevector states accessible near the $\Gamma$ point due to Bragg scattering induced by the periodicity of the crystal lattice. They manifest as a dip in planar reflectance and are usually phenomenologically described by adding oscillators to the material’s transverse dielectric function  \cite{Bluet1999, Nakashima1997}. The negative dispersion of the LO phonon ensures that these weak phonon modes exist within the Reststrahlen band, co-existing in frequency with propagating or localised SPhPs. This is illustrated in Fig.~\ref{fig:Fig1}a for 4H-SiC, the material studied in this Letter, whose weak phonon lies at around $837.5/$cm. Around 250 unique polytypes of SiC exist, each with different weak phonon frequencies, allowing the weak LO phonon to be tuned throughout the Reststrahlen region. The weak phonons of 15R- and 6H-SiC for example lie near $860$ and $885$/cm respectively.\\
Longitudinal-transverse hybridisation has been also theoretically predicted in polar quantum wells and superlattices \cite{Ridley1992,Chen2004}, and realised in plasmonic systems, where nanoscale confinement of transverse plasmonic modes makes very large wavevectors accessible, intersecting the negative dispersion of the longitudinal oscillation of the electron gas \cite{Ciraci2012, Ciraci2013}. This results in red shift of the modal frequency, which can become non-negligible when an appreciable portion of the plasmonic field exists at large wavevectors.  
Contrastingly in the systems under investigation Bragg folding ensures that the longitudinal mode is accessible for all values of the wavevector, which as we will show leads to a strong hybridisation even in optically large resonators. \\
In the following we will initially develop a microscopic theory of light-matter coupling in polar dielectric systems including spatial dispersion. This theory will be then used to theoretically investigate the reflectance of a 4H-SiC surface. Finally,  we will present experimental results demonstrating strong coupling, and thus the existence of LTPP, using arrays of 4H-SiC nanopillars. 

\section{Theory}
Our starting point in order to microscopically model the hybridisation of phonon polaritons with ZFLOs is to expand the theory describing ionic motion in a polar dielectric \cite{Trallero-Giner1992,Trallero-Giner1994a,Santiago-Perez2014} to the retarded regime.
In frequency-space the material displacement $\mathbf{X}$ obeys the equation
\begin{align}
\nonumber \left[\omega_{\mathrm{T}}^2-\omega(\omega+i\gamma)\right] \mathbf{X} &= -\beta_{\mathrm{L}}^2 \nabla(\nabla \cdot \mathbf{X})+\beta_{\mathrm{T}}^2 \nabla \times\nabla\times \mathbf{X} \\&\quad -\frac{\alpha}{\rho} (\nabla \phi-i\omega \mathbf{A}),\label{eq:eom}
\end{align}
where $\phi (\mathbf{A})$, is the electromagnetic scalar (vector) potential, the material high-frequency dielectric constant is $\epsilon_{\infty}$,  the transverse (longitudinal) optical phonon frequency at the $\Gamma$ point is $\omega_{\mathrm{T}} (\omega_{\mathrm{L}})$, the material density is given by $\rho$, the phonon damping rate by $\gamma$, and the transverse (longitudinal) phonon velocities in the limit of quadratic dispersion by $\beta_{\mathrm{T}} (\beta_{\mathrm{L}})$. Finally the light matter coupling strength is given by the polarizability
\begin{align}
	\alpha = \sqrt{\rho \epsilon_0 \epsilon_{\infty} \left(\omega_{\mathrm{L}}^2 - \omega_{\mathrm{T}}^2\right)},
\end{align}
and we assume that the only effect of the anisotropy is the Bragg folding along the c-axis.  \\
In the Supplemental Information this equation of motion, in conjunction with Maxwell equations and the appropriate electromagnetic and mechanical boundary conditions, are solved by the introduction of auxiliary scalar and vector potentials $\mathrm{Y} = \nabla \cdot \mathbf{X},\; \boldsymbol{\Gamma} = \nabla \times \mathbf{X}$, allowing us to write the ionic displacement as
\begin{align}
\nonumber
\mathbf{X} = \frac{1}{\omega_{\mathrm{T}}^2-\omega(\omega+i\gamma) } &\left[\beta_{\mathrm{T}}^2 \nabla \times \boldsymbol{\Gamma} - \frac{ \beta_{\mathrm{L}}^2 \epsilon_{\infty}}{\epsilon\left(\omega, 0\right)} \nabla \mathrm{Y} \right.\\&\left. - \frac{\alpha}{\rho} (\nabla  \phi_{\mathrm{H}}-i\omega \mathbf{A}) \right], 
\label{eq:recons}
\end{align}
where $\phi_{\mathrm{H}}$ is the homogeneous electric scalar potential,  solution to the Laplace equation, and $\epsilon\left(\omega,0\right)$ is the lattice dielectric function in the absence of spatial dispersion. \\

\subsection*{Application to Surface Phonon Polaritons}
\begin{figure}
	\centering
	\includegraphics[width=\columnwidth]{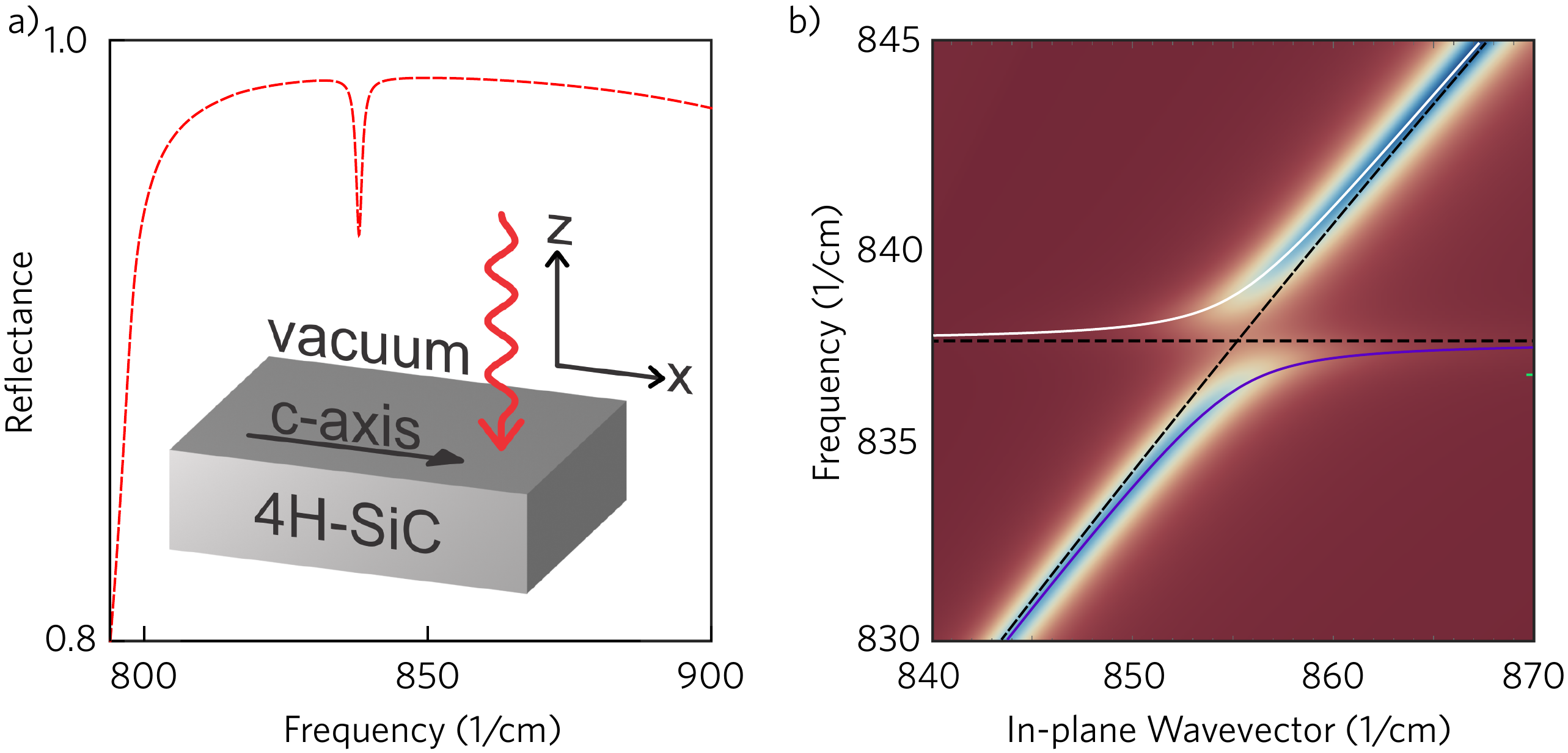}
	\caption{\label{fig:Fig2} a) TM polarised reflectance calculated for an a-cut, 4H-SiC substrate supporting a weak phonon mode at near normal incidence. The in-plane wavevector of the incident light is parallel to the crystal c-axis. The inset illustrates the physical system under study. b) Colormap illustrates the dispersion of the surface phonon polariton on the bilayer interface calculated by Eq.~\ref{eq:theref}. Overlaid dashed lines indicate the bare weak phonon and surface phonon polariton dispersions, while the white and purple curves illustrate the coupled LTPP modes.}
\end{figure}
In order to clearly demonstrate how our theory leads to the appearance of hybrid LTPP modes, here we apply it to the analytically solvable case of an a-cut uniaxial polar dielectric halfspace in vacuum, shown in the inset of Fig.~\ref{fig:Fig2}a. We choose an a-cut crystal because in this system the ZFLO phonon manifests as a dip in the planar reflectance, permitting us to compare our analytical solution with experimental observation of ZFLOs previously reported in the literature \cite{Nakashima1997, Caldwell2013, Bluet1999, Ellis2016}. As derived in the Supplemental, the Fresnel coefficient for TM polarised light incident along the c-axis, including spatial dispersion, is
\begin{equation}
	r= \frac{k_{\mathrm{B}} - \frac{k_{\mathrm{T}}}{\epsilon\left(\omega,k\right)} -  \Omega}{k_{\mathrm{B}}  + \frac{k_{\mathrm{B}}}{\epsilon\left(\omega,k\right)} + \Omega},
	\label{eq:theref}
\end{equation}
where $k_{\mathrm{B}} (k_{\mathrm{T}})$ are the out-of-plane wavevector components of the transverse mode in the vacuum (dielectric), $\epsilon\left(\omega,k\right)$ is the dielectric function parallel to the c-axis including spatial dispersion, and $\Omega$ encodes the mechanical boundary condition $ \bar{\sigma} \cdot \mathbf{z}\rvert_{z=0} = 0$, where $\bar{\sigma}$ is the stress tensor. We can apply this result to the description of the ZFLO modes observed in planar reflectance measurements of a-cut SiC polytypes by shifting the longitudinal in-plane wavevector to
\begin{equation}
	k_x \to  \frac{2 \pi}{c} + k_x,
\end{equation}
where $c$ is the length of the unit cell along the c-axis and $k_x$ is the in-plane wavevector of the incident photons. The result for a 4H-SiC substrate is shown in Fig~\ref{fig:Fig2}a, where the anisotropy is reintroduced by considering the transverse wavevector to be that of the extraordinary wave in the crystal, yielding a characteristic dip in the reflectance at the ZFLO frequency $837.5/$cm, consistent with previously reported experimental data \cite{Bluet1999, Ellis2016}. \\
This result also allows for investigation of the guided modes of the planar structure, satisfying
\begin{equation}
	k_{\mathrm{B}}  + \frac{k_{\mathrm{T}}}{\epsilon\left(\omega,k\right)} + \Omega = 0,
\end{equation}
whose dispersion is seen in Fig.~\ref{fig:Fig2}b, where the imaginary component of the reflectance coefficient Eq.~\ref{eq:theref} is plotted utilising standard parameters for the 4H-SiC dielectric function with damping rate $\gamma = 4/$cm. The clearly visible anticrossing shows that the SPhP supported on the planar interface strongly couples with the ZFLO phonon, leading to the hybrid longitudinal-transverse excitations we named LTPP.

\section{Experimental Results}
In order to verify the existence of LTPP, we consider square arrays of cylindrical 4H-SiC resonators on a same-material c-cut substrate \cite{Caldwell2013, Gubbin2016, Chen2014}. Such systems, sketched in Fig.~\ref{fig:Fig1}b,  support a variety of transverse SPhP modes, with highly tuneable frequencies dependent on the geometrical parameters \cite{Gubbin2016,Gubbin2017}. The monopole mode in particular, polarised out-of the substrate plane, is highly sensitive to the interpillar spacing (pitch) due to interpillar dipole-dipole coupling and can effectively be tuned throughout the Reststrahlen band \cite{Chen2014}. This mode is often referred to as longitudinal in the literature but this naming convention only refers to the electric field orientation with respect to the pillar long axis, the mode is nonetheless electromagnetically transverse, with non-vanishing curl. Apart from its technological relevance thanks to small mode volumes and narrow and tuneable resonances, this system presents a key advantage over the planar system discussed in the previous section. The SPhPs here exist within the light-line, and it is thus possible to spectroscopically probe the anti-crossing without the need for complex prism coupling set-ups \cite{Passler2017}. \\
The underlying mechanism that gives rise to the longitudinal-transverse hybridization mode in the theoretical treatment and experimental results is the same. However, the nanostructured array complicates the theoretical analysis since there is no analytical solution to Eq.~\ref{eq:eom} for this system. The experimental practicality of the nano pillar array comes in fact at the cost of heavy numerical complications, which dramatically increase the computational power required to solve the problem. This renders the full 3D calculations necessary to describe the resonator array on a substrate is prohibitively expensive.\\
\begin{figure}
	\centering
	\includegraphics[width=\columnwidth]{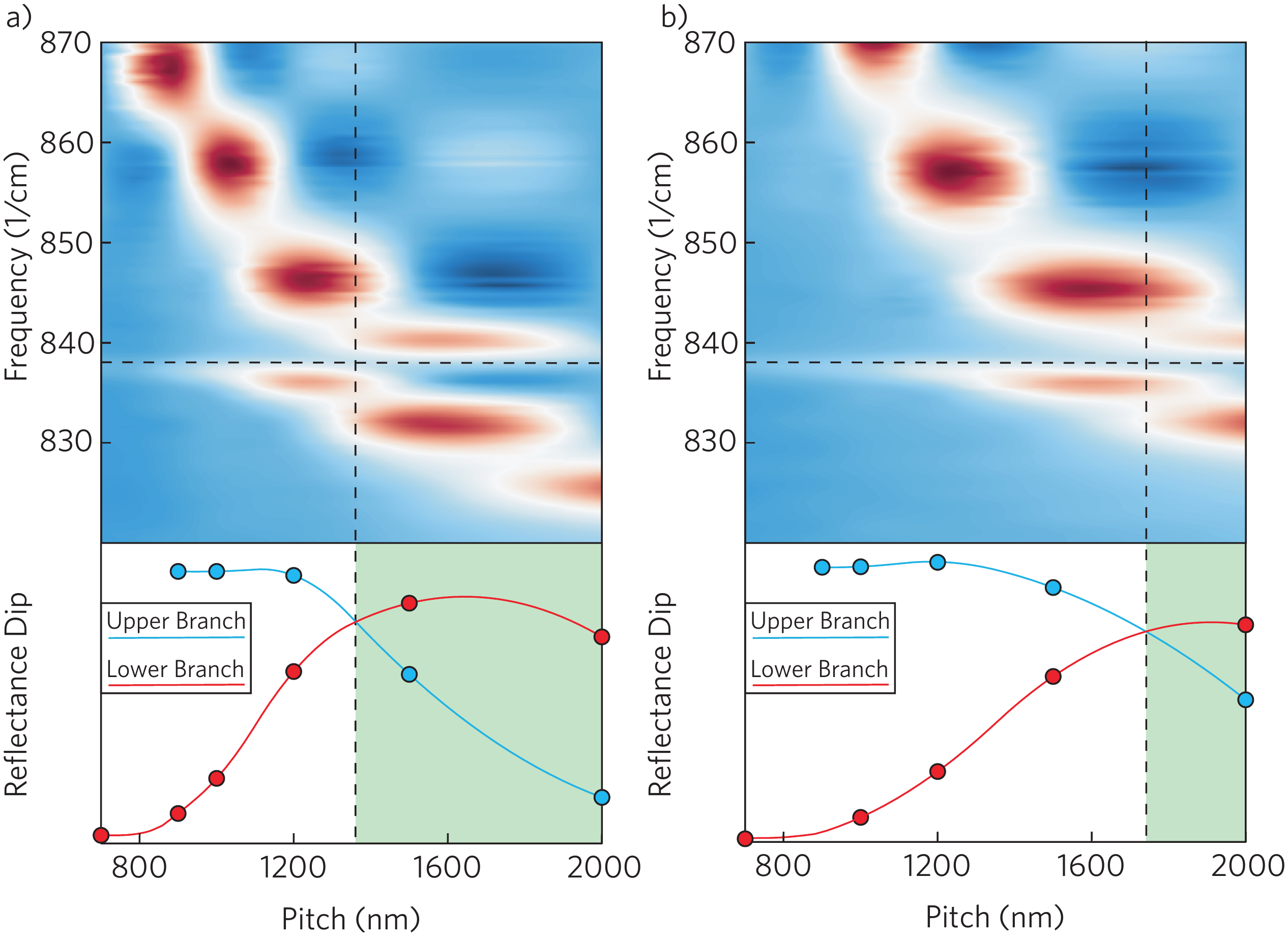}
	\caption{\label{fig:Fig3} Upper panels illustrate the experimental reflectance from an array of 4H-SiC nanopillars with diameter a) 300 nm b) 500 nm and height 950 nm recorded as a function of the lattice period. The horizontal dashed line indicates the weak phonon resonance. The lower panel shows the magnitude of the reflectance dip extracted from the data. The shaded region indicates the region where the upper polariton is predominantly LO in character.}
\end{figure}
We exploit the broad tunability of the monopole mode by fabricating samples with interpillar spacings in the range $700-2000$ nm, over which the resonance is expected to tune from the high to the low energy side of the ZFLO phonon at $837.5/$cm. Pillars were fabricated with a uniform height of $950$ nm and diameters of $300$ and $500$ nm. Nanopillar arrays were fabricated from semi-insulating c-cut 4H-SiC substrates by reactive ion etching \cite{Caldwell2013}. We probe the planar reflectance utilising Fourier transform infrared spectroscopy.\\
Results are shown in Fig.~\ref{fig:Fig3} for pillars of nominal diameters $300$ nm (a) and $500$ nm (b) for the full range of interpillar spacings explored. A larger diameter results in a blue shift of the monopolar mode  \cite{Gubbin2017}, while increasing the interpillar spacing causes the monopolar mode to red shift as a result of decreased coupling between resonators. When the monopolar mode approaches the 4H-SiC ZFLO, illustrated by the horizontal dashed line in Fig.~\ref{fig:Fig3}, a second branch appears in the reflectance on the low energy side of the ZFLO mode. Rather than continuing to red shift through the ZFLO for large interpillar spacings, the monopolar mode remains on it's high energy side and the new branch red shifts. This anti-crossing behaviour, previously illustrated in Sec. 1, is a hallmark of strong coupling and demonstrates that the monopolar phonon polariton mode of the pillar array is hybridised with the ZFLO phonon \cite{Gubbin2016,Passler2018}. \\
Further evidence for the hybrid nature of the observed LTPP resonances can be acquired from the magnitude of the recorded reflectance dips, calculated by subtracting the reflectance of the pillar array from that of the planar substrate, as shown in the lower panel of Fig.~\ref{fig:Fig3}. In this panel the blue (red) circles correspond to the upper (lower) branches in the upper panel. For small or large interpillar spacings, where the detuning between monopolar and ZFLO modes is large we see that, as expected, the upper and lower branches have characteristics of the bare modes. The monopolar mode of the resonator array couples well to the impinging light resulting in a deep reflectance dip, while the ZFLO couples weakly. In the intermediate region instead, the modes are linear combinations of the bare monopolar and ZFLO modes, leading to the crossing for the reflectance dip which demonstrates their hybrid nature. In the lower panel of Fig.~\ref{fig:Fig3} the shaded regions indicate where the upper (lower) branch is more ZFLO (monopolar) in character. Particularly the equalisation of the branches reflectance is a signal of the anticrossing point, indicating that both branches are composed of equal parts monopolar and ZFLO modes. This can be seen by following the vertical lines up onto the reflectance maps, corresponding to the avoided crossing of the LTPP branches. \\
Additional samples were fabricated with the same nominal array parameters. Wide inter-sample variability ensures that these arrays have different monopolar frequencies. All samples reproduce the anticrossing at the ZFLO frequency, this data is available in the Supplemental.
These results illustrate the hybridisation of longitudinal and transverse modes in polar dielectric structures, thus providing the first clear experimental evidence of the LTPP modes theoretically predicted in the previous Section. This hybridisation, mediated by the mechanical boundary conditions at the crystal surface, cannot be achieved in bulk as the longitudinal mode cannot be matched without an interface. This kind of surface-induced hybridisation is well understood in plasmonic systems, where spatial dispersion arises as a result of electron pressure \cite{Ciraci2013}. In plasmonic systems however these effects are only accessible where the electric field is confined on the nanoscale, meaning that resonances are comprised of sufficiently high wavevector Fourier components to experience the dispersion \cite{Ciraci2012}. In the polar dielectric systems discussed here these large wavevector Fourier components are instead accessible in optically large resonators meaning that the hybridisation is essentially accessible in any appropriately tuned polar dielectric resonator.

\section{Conclusions}
Fabrication of resonators whose eigenmodes are linear superpositions of transverse and longitudinal waves has important technological implications. Such modes can be directly pumped electrically through the Fr{\"o}hlich interaction, allowing for the creation of efficient  electroluminescent devices operating throughout the SiC Reststrahlen band. An efficient injection scheme could also potentially lead to the development of coherent phonon polariton-based light sources, an idea which has received some attention in recent literature \cite{OhtaniX,CartellaX}.
Further flexibility can be found by applying these results to superlattice systems in which the Brillouin folding can be finely tuned \cite{Colvard1985} and the hybrid material dielectric function can be controlled \cite{Caldwell2016}, potentially allowing for the creation of electroluminescent devices operating across the mid-infrared spectral region.

\section*{Funding Information}
S.D.L. is a Royal Society Research Fellow. S.D.L and C.R.G. acknowledge support from the Innovation Fund of the EPSRC Programme EP/M009122/1. C.T.E. and J.G.T.  acknowledge support from the Office of Naval Research. M.A.M. and C.T.E. acknowledge support from the National Research Council Research Associateship program. R.B. acknowledges the Capes Foundation for a Science Without Borders fellowship (Bolsista da Capes, Proc. No. BEX 13.298/13-5).

\section*{Supplemental Documents}
See the Supplementary Information for supporting content.


\bibliographystyle{apsrev4-1}
\bibliography{sampleup}

\pagebreak
\begin{center}
\textbf{\large Supplementary Information}
\end{center}
\setcounter{equation}{0}
\setcounter{section}{0}
\setcounter{figure}{0}
\setcounter{table}{0}
\setcounter{page}{1}
\renewcommand{\theequation}{S\arabic{equation}}
\renewcommand{\thefigure}{S\arabic{figure}}
\renewcommand{\bibnumfmt}[1]{[S#1]}
\renewcommand{\citenumfont}[1]{S#1}

\section{Formalism}

We can write the ionic equation of motion for a polar dielectric crystal coupled to an electric field, in the form 
\begin{align}
\nonumber \left[\omega_{\mathrm{T}}^2-\omega(\omega+i\gamma)\right] \mathbf{X} &= -\beta_{\mathrm{L}}^2 \nabla(\nabla \cdot \mathbf{X})+\beta_{\mathrm{T}}^2 \nabla \times\nabla\times \mathbf{X} \\&\quad -\frac{\alpha}{\rho} (\nabla \phi - i \omega \mathbf{A}),\label{eq:eom}
\end{align}
where $\omega_{\mathrm{T}}$ is the polar dielectric's transverse phonon frequency, $\gamma$ the phonon damping rate, $\mathbf{X}$ is the ionic displacement, $\alpha$ is the light-matter coupling strength, $\rho$ the density, $\beta_{\mathrm{T}} \;(\beta_{\mathrm{L}})$ the transverse (longitudinal) phonon velocities and $\phi \; (\mathbf{A})$ is the scalar (vector) potential related to the electromagnetic fields $\mathbf{E}, \mathbf{H}$ by
\begin{align}
	\mathbf{E} &= - \nabla \phi + i \omega \mathbf{A},\nonumber\\
	\mathbf{H} &= \frac{1}{\mu_0} \nabla \times \mathbf{A}.
\end{align}\\
Taking the divergence (curl) of Eq.~\ref{eq:eom} and defining auxiliary scalar (vector) potentials $\mathrm{Y} = \nabla \cdot \mathbf{X}, \;(\boldsymbol{\Gamma} = \nabla \times \mathbf{X})$ we can find decoupled equations of motion for the longitudinal auxiliary potential
\begin{align}
\left[(\omega_{\mathrm{L}}^2-\omega(\omega+i\gamma))+\beta_{\mathrm{L}}^2\nabla^2\right] \mathrm{Y}&=0, \label{eq:longeom}
\end{align}
and transverse one
\begin{align}
\left[(\omega_{\mathrm{T}}^2-\omega(\omega+i\gamma))+\beta_{\mathrm{T}}^2\nabla^2\right]\boldsymbol{\Gamma} &=i \omega \frac{\alpha}{\rho} \nabla\times\mathbf{A}\label{eq:transeom},
\end{align}
where we utilised the Coulomb gauge condition $\nabla \cdot \mathbf{A} = 0$,  the absence of free charges $\nabla\cdot \mathbf{D} = 0$, and the constitutive relation linking the electric and displacement fields to the ionic displacement
\begin{equation}
	\mathbf{D} = \epsilon_0 \epsilon_{\infty} \mathbf{E} + \alpha \mathbf{X}.
	\label{eq:Const}
\end{equation}
We also introduced the longitudinal phonon frequency
\begin{equation}
	\omega_{\mathrm{L}}^2 = \omega_{\mathrm{T}}^2 + \frac{\alpha^2}{\epsilon_0 \epsilon_{\infty} \rho},
\end{equation}
where $\epsilon_{\infty}$ is the high-frequency dielectric constant of the polar dielectric. In terms of the newly defined potentials we can reconstruct the ionic displacement by substitution as
\begin{align}
\mathbf{X} = \frac{1}{\omega_{\mathrm{T}}^2-\omega(\omega+i\gamma) } \biggr[&\beta_{\mathrm{T}}^2 \nabla \times \boldsymbol{\Gamma} - \frac{ \beta_{\mathrm{L}}^2 \epsilon_{\infty}}{\epsilon\left(\omega, 0\right)} \nabla \mathrm{Y} \nonumber \\
	&- \frac{\alpha}{\rho} (\nabla  \phi_{\mathrm{H}}- i \omega \mathbf{A}) \biggr], \label{eq:recons}
\end{align}
where we recognised the dielectric function of a polar dielectric crystal in the absence of spatial dispersion
\begin{equation}
	\epsilon\left(\omega, 0\right) = \epsilon_{\infty} \frac{\omega_{\mathrm{L}}^2 - \omega\left(\omega + i \gamma\right)}{\omega_{\mathrm{T}}^2 - \omega \left(\omega + i \gamma\right)}. \label{eq:StaticDie}
\end{equation}

\subsection{Longitudinal Equation}
The longitudinal equation of motion Eq.~\ref{eq:longeom} is just a scalar Helmholtz equation. Taking the divergence of the displacement field $\mathbf{D}$ as defined in Eq.~\ref{eq:Const} we can find
\begin{equation}
	\nabla^2 \phi = \frac{\alpha}{\epsilon_0 \epsilon_{\infty}} \mathrm{Y} \label{eq:QuasiConst},
\end{equation}
whose solution is simply given by
\begin{equation}
\phi = \phi_{\mathrm{H}} -\frac{\alpha}{\epsilon_0\epsilon_{\infty}}\frac{\beta_{\mathrm{L}}^2}{\omega_{\mathrm{L}}^2-\omega(\omega+i\gamma)}\mathrm{Y} \label{eq:Scapot},
\end{equation}
where $\phi_{\mathrm{H}}$ is the homogeneous solution of Eq.~\ref{eq:QuasiConst} satisfying $\nabla^2 \phi_{\mathrm{H}} = 0$.

\subsection{Transverse Equation}
Using Maxwell's curl equation in conjunction with the constitutive relation in Eq.~\ref{eq:Const}
\begin{align}
	\nabla \times \mathbf{H} = \frac{\partial \mathbf{D}}{\partial t} = \epsilon_0 \epsilon_{\infty} \frac{\partial \mathbf{E}}{\partial t} + \alpha \frac{\partial \mathbf{X}}{\partial t},
\end{align}
we can put Eq.~\ref{eq:transeom} for the transverse potential in the form
\begin{align}
	i \mu_0 \alpha \omega \boldsymbol{\Gamma} &= \left[ \frac{\omega^2 \epsilon_{\infty}}{c^2} + \nabla^2 \right] \nabla \times \mathbf{A}.
\end{align}
Making a spatial Fourier transform $\nabla^2 \to - k^2$ and substituting this back into the transverse equation of motion we arrive at the result
\begin{equation}
	\left[k^2 - \frac{\omega^2}{c^2} \epsilon\left(\omega,k\right) \right]\boldsymbol{\Gamma} = 0, \label{eq:transeom2}
\end{equation}
where the easily identified spatially dispersive dielectric function is given by
\begin{equation}
	\epsilon\left(\omega,k\right) = \epsilon_{\infty} \frac{\omega_{\mathrm{L}}^2 - \omega \left(\omega + i \gamma\right) - \beta_{\mathrm{T}}^2 k^2}{\omega_{\mathrm{T}}^2 - \omega \left(\omega + i \gamma\right) - \beta_{\mathrm{T}}^2 k^2},
\end{equation}
which reduces to the non-dispersive Eq.~\ref{eq:StaticDie} in the limit $\beta_{\mathrm{T}} k \ll \omega$.

\section{Application to a Polar Halfspace}
Here we apply the general theory previously derived to the specific case of an a-cut polar dielectric halfspace occupying $z<0$ whose c-axis is aligned with the x-axis. The region $z>0$ is filled with a non-resonant dielectric whose dispersionless dielectric function is given by $\epsilon_{\mathrm{B}}$. The system is illuminated from $z=\infty$ by a TM polarised electromagnetic field with incident wavevector in the $xz$ plane.\\
As we saw in the general theory, solutions to the general isotropic equation of motion in the lower halfspace Eq.~\ref{eq:eom} are separable into three classes. Firstly the homogeneous solution, satisfying $\nabla^2 \phi_{\mathrm{H}} = 0$ whose solution is given by
\begin{equation}
	\phi_{\mathrm{H}} = - \frac{t_{\mathrm{H}}}{\epsilon_0 \omega} e^{i k_x x} e^{k_x z},	
\end{equation}
yields the following electric and displacement fields in the lower halfspace
\begin{align}
	\mathbf{E}_{\mathrm{H}} &= - \nabla \phi_{\mathrm{H}} =  \frac{t_{\mathrm{H}} k_x}{\omega \epsilon_0} (i \hat{\mathbf{x}}+ \hat{\mathbf{z}})e^{ik_x x+k_x z},\nonumber \\
	\mathbf{X}_{\mathrm{H}} &=\frac{\alpha}{\rho\epsilon_0 \omega} \frac{ t_{\mathrm{H}} k_x}{\omega_{\mathrm{T}}^2 - \omega \left(\omega + i \gamma\right)}  (i \hat{\mathbf{x}}+ \hat{\mathbf{z}})e^{ik_x x+k_x z},
\end{align}
which are chosen to ensure decay away from the interface as $z \to - \infty$. Here $t_{\mathrm{H}}$ is a constant to be determined by application of the electromagnetic and mechanical boundary conditions. In the upper halfspace where $\alpha = 0$ it is clear that the scalar equation Eq.~\ref{eq:QuasiConst} still admits a homogeneous solution whose out-of-plane wavevector is equal and opposite to that in the lower halfspace. Considering this mode allows the homogeneous solution to be ignored in calculation of the electromagnetic boundary conditions.

Zone-centre longitudinal modes are not considered as solutions of Eq.~\ref{eq:longeom} due to their far off-resonant nature. As a result of Bragg scattering along the crystal c-axis, which for the a-cut system considered is parallel to the x-axis, the in-plane wavevector of the Bragg scattered modes is shifted as $k_x \to k_{\mathrm{M}} + k_x$ in which $k_{\mathrm{M}} = 2 \pi/a$ where $a$ is the lattice constant along the c-axis. The scalar potential is therefore given by
\begin{equation}
	\mathrm{Y} = \frac{\alpha \mu_0 t_{\mathrm{L}}}{\rho \omega} e^{i \left(k_{\mathrm{M}} + k_x\right) x+ik_\mathrm{L} z},
\end{equation}
in which $t_{\mathrm{L}}$ is a constant to be determined by application of the appropriate boundary conditions. These modes are subject to the dispersion relation
\begin{equation}
	0=\omega_{\mathrm{L}}^2-\omega(\omega+i\gamma)-\beta_{\mathrm{L}}^2(k_{\mathrm{M}}+k_x)^2-\beta_{\mathrm{L}}^2  k_{\mathrm{L}}^2,
\end{equation}
which yields the out-of-plane longitudinal wavevector $k_{\mathrm{L}}$.\\
The corresponding electric potential is calculated through the inhomogeneous component of Eq.~\ref{eq:Scapot} as
\begin{align}
	\phi_{\mathrm{L}} &= - \frac{\alpha^2}{\rho \epsilon_0^2 \epsilon_{\infty} \omega} \frac{\beta_{\mathrm{L}}^2/c^2}{\omega_{\mathrm{L}}^2 - \omega \left(\omega + i \gamma\right)} t_{\mathrm{L}} e^{i \left(k_{\mathrm{M}} + k_x\right)  x} e^{i k_{\mathrm{L}} z},
\end{align}
generating the electric field
\begin{align}
	\mathbf{E}_{\mathrm{L}} = & - \nabla \phi_{\mathrm{L}}, \nonumber 
	 =  \frac{i \beta_{\mathrm{L}}^2 t_{\mathrm{L}}}{c^2 \epsilon_0 \omega}   \left[1 - \frac{\epsilon_{\infty}}{\epsilon\left(\omega,0\right)}\right] \nonumber \\
	& \times \left[ \left(k_{\mathrm{M}} + k_x\right) \hat{\mathbf{x}} +  k_{\mathrm{L}} \hat{\mathbf{z}}\right] e^{i \left(k_{\mathrm{M}} + k_x\right)  x} e^{i k_{\mathrm{L}} z},
\end{align}
and ionic displacement through Eq.~\ref{eq:recons}
\begin{align}
\mathbf{X}_{\mathrm{L}} =& - \frac{\beta_{\mathrm{L}}^2 }{\omega_{\mathrm{L}}^2-\omega(\omega+i\gamma) } \nabla \mathrm{Y}\nonumber 
	= -  \frac{i \alpha t_{\mathrm{L}}}{\rho \epsilon_0 \omega}  \frac{\beta_{\mathrm{L}}^2 / c^2}{\omega_{\mathrm{L}}^2-\omega(\omega+i\gamma) } \nonumber \\
	& \times \left[ \left(k_{\mathrm{M}} + k_x\right) \hat{\mathbf{x}} +  k_{\mathrm{L}} \hat{\mathbf{z}}\right] e^{i \left(k_{\mathrm{M}} + k_x\right)  x+ik_\mathrm{L} z}.
\end{align}\\
Finally as a solution to Eq.~\ref{eq:transeom} we consider a TM polarised transverse field, whose magnetic component is given by
\begin{align}
	\mathbf{H} = t_{\mathrm{T}}  e^{i k_x x} e^{i k_{\mathrm{T}} z} \hat{\mathbf{y}},
\end{align}
where $t_{\mathrm{T}}$ is a constant to be determined by applying the electromagnetic and mechanical boundary conditions and $k_{\mathrm{T}}$ is the out-of-plane wavevector for the transverse mode. This mode is generated by the vector potential
\begin{align}
	\mathbf{A} &= \frac{t_{\mathrm{T}}}{\omega^2 \epsilon_0 \epsilon\left(\omega, k\right)} \left[- i k_{\mathrm{T}} \hat{\mathbf{x}} + i k_x \hat{\mathbf{z}}\right] e^{i k_x x} e^{i k_{\mathrm{T}} z},
\end{align}
from which we can calculate the auxiliary vector potential
\begin{align}
	\boldsymbol{\Gamma} &= \frac{i}{\alpha \omega} \frac{\omega^2}{c^2} \left[ \epsilon\left(\omega, k\right) - \epsilon_{\infty} \right]   t_{\mathrm{T}} e^{i k_x x} e^{i k_{\mathrm{T}} z} \hat{\mathbf{y}}.
\end{align}
Now we are in a position to calculate the transverse electric field
\begin{align}
	\mathbf{E}_{\mathrm{T}} &= i\omega \mathbf{A} =  \frac{t_{\mathrm{T}}}{\omega \epsilon_0 \epsilon\left(\omega, k\right)} \left[k_{\mathrm{T}} \hat{\mathbf{x}} -  k_x \hat{\mathbf{z}}\right] e^{i k_x x} e^{i k_{\mathrm{T}} z},
\end{align}\\
and the transverse component of the material displacement through Eq.~\ref{eq:recons}
\begin{align}
\mathbf{X}_{\mathrm{T}} =& \frac{1}{\omega_{\mathrm{T}}^2-\omega(\omega+i\gamma) } \left[\beta_{\mathrm{T}}^2 \nabla \times \boldsymbol{\Gamma} + i \omega \frac{\alpha}{\rho} \mathbf{A}\right] ,\nonumber \\
=& \frac{\alpha}{\rho \epsilon_0 \omega} \frac{ t_{\mathrm{T}}  \left(k_{\mathrm{T}} \hat{\mathbf{x}} - k_x \hat{\mathbf{z}}\right)}{\omega_{\mathrm{T}}^2-\omega(\omega+i\gamma) } \left[ \frac{1}{\epsilon\left(\omega, k\right)} \right.\\&\left.+\frac{\beta_{\mathrm{T}}^2 \omega^2}{c^2} \frac{1}{\omega_{\mathrm{T}}^2 - \omega\left(\omega+ i \gamma\right) - \beta_{\mathrm{T}}^2 k^2} \right] \nonumber e^{i k_x x} e^{i k_{\mathrm{T}} z}.
\end{align}\\
In the above the out-plane-wavevector for the transverse mode $k_T$ is given by the root of Eq.~\ref{eq:transeom2}. In practice as the wavevector of the transverse mode is small, satisfying $k \beta_{\mathrm{T}} \ll \omega$ it is only necessary to consider this equation in the non-dispersive limit where it reduces to the standard Helmholtz equation.\\
The electromagnetic fields in the upper halfspace $z>0$ correspond to a TM polarised plane wave incident from $z = \infty$. They are given by
\begin{align}
\mathbf{E}_{\mathrm{I}} &= \frac{e^{i k_x x}}{\epsilon_0 \epsilon_\mathrm{B} \omega}  \left[ \left( k_{\mathrm{B}} \hat{\mathbf{x}} - k_x \hat{\mathbf{z}}\right) e^{i k_{\mathrm{B}} z}  - r \left(  k_{\mathrm{B}} \hat{\mathbf{x}} + k_x \hat{\mathbf{z}}\right)e^{- i k_{\mathrm{B}} z} \right],\nonumber\\
	\mathbf{H}_{\mathrm{I}} &= \left[e^{i k_{\mathrm{B}} z} + r e^{- i k_{\mathrm{B}} z} \right] e^{i k_x x},
\end{align}
where $r$ is the reflection coefficient and $k_{\mathrm{B}}$ is the out-of-plane wavevector in the upper halfspace given by
\begin{align}
	k_{\mathrm{B}} = \sqrt{\epsilon_{\mathrm{B}} \frac{\omega^2}{c^2} - k_x^2}.
\end{align}

\subsection{Boundary Conditions}
We start by applying the standard electromagnetic Maxwell boundary conditions. Continuity of the tangential magnetic field yields
\begin{equation}
	1 + r = t_{\mathrm{T}},
\end{equation}
and that of the tangential electric field gives
\begin{align}
	\frac{k_{\mathrm{B}}}{\epsilon_{\mathrm{B}} }\left(1 - r\right) &= \frac{k_{\mathrm{T}} }s{  \epsilon\left(\omega, k\right)} t_{\mathrm{T}} \nonumber \\
	&\quad +  i \left(k_{\mathrm{M}}+ k_x\right)\frac{\beta_{\mathrm{L}}^2}{c^2}  \left(1 - \frac{\epsilon_{\infty}}{\epsilon\left(\omega,0\right)}\right)t_{\mathrm{L}},
\end{align}
where we eliminate the homogeneous electric field by considering in addition its counterpart in the upper halfspace as previously described.\\
To account for the oscillations of the crystal we also need to apply mechanical boundary conditions. The appropriate choice for a free surface such as that considered here are the continuity of the mechanical forces, or the normal components of the stress tensor. These boundary conditions can be written in the form
\begin{align}
	\frac{\partial \mathrm{X}_x}{\partial z} + \frac{\partial \mathrm{X}_z}{\partial x} &= 0,\nonumber\\
	C_{13} \frac{\partial \mathrm{X}_x}{\partial x} + C_{33} \frac{\partial \mathrm{X}_z}{\partial z} &= 0,
\end{align}
where $C_{13},\; C_{33}$ are elastic coefficients of the lattice. Application of the former boundary condition yields
\begin{align}
	t_{\mathrm{H}} =& \frac{1}{k_x^2}\biggr[i   \frac{\beta_{\mathrm{L}}^2}{ c^2} \frac{\epsilon_{\infty}}{\epsilon\left(\omega,0\right)}  t_{\mathrm{L}} \left(k_{\mathrm{M}} + k_x\right)k_{\mathrm{L}} -\frac{t_{\mathrm{T}}}{2} \left(k_{\mathrm{T}}^2  - k_x^2\right)  \\
	& \times \left(\frac{\beta_{\mathrm{T}}^2 }{c^2} \frac{\omega^2}{\omega_{\mathrm{T}}^2 - \omega\left(\omega+ i \gamma\right) - \beta_{\mathrm{T}}^2 k^2} +  \frac{1}{\epsilon\left(\omega, k\right)}\right)  \biggr],\nonumber
\end{align}
and combining with the latter condition we can find
\begin{align}
	t_{\mathrm{L}} &= \frac{t_{\mathrm{T}}}{2}\frac{c^2}{\beta_{\mathrm{L}}^2} \frac{\epsilon\left(\omega,0\right)}{\epsilon_{\infty}} \nonumber \\
	&\quad \times \frac{\left(C_{33} - C_{13}\right) \left(2 i k_x k_{\mathrm{T}} +  \left(k_{\mathrm{T}}^2 - k_x^2\right) \right) }{C_{13} (k_{\mathrm{M}} + k_x)^2 + C_{33} k_{\mathrm{L}}^2 + i \left(C_{33} - C_{13}\right) \left(k_{\mathrm{M}} + k_x\right) k_{\mathrm{L}}} \nonumber \\
	&\quad \times \left[\frac{\beta_{\mathrm{T}}^2 }{c^2} \frac{\omega^2}{\omega_{\mathrm{T}}^2 - \omega\left(\omega+ i \gamma\right) - \beta_{\mathrm{T}}^2 k^2} +  \frac{1}{\epsilon\left(\omega, k\right)}\right].
\end{align}\\
Finally, combining this result with the electromagnetic boundary conditions we can find a single equation whose solution yields the reflectance coefficient of the halfspace.
\begin{align}
	\frac{k_{\mathrm{B}}}{\epsilon_{\mathrm{B}}}\left(1 - r\right) =\left(1+r\right) \left[ \frac{k_{\mathrm{T}}}{\epsilon\left(\omega, k\right)}  + \Omega \right],
\end{align}
where
\begin{align}
	 \Omega & = \frac{i \left(k_{\mathrm{M}} + k_x\right)}{2}   \left(\frac{\epsilon\left(\omega,0\right)}{\epsilon_{\infty}} - 1\right) \nonumber \\
	 &\quad \times \frac{\left(C_{33} - C_{13}\right) \left(2 i k_x k_{\mathrm{T}} +  \left(k_{\mathrm{T}}^2 - k_x^2\right) \right) }{C_{13} \left(k_{\mathrm{M}} + k_x\right)^2 + C_{33} k_{\mathrm{L}}^2 + i \left(C_{33} - C_{13}\right) \left(k_{\mathrm{M}} + k_x\right) k_{\mathrm{L}}} \nonumber \\
	 &\quad \times  \left[\frac{\beta_{\mathrm{T}}^2 \omega^2}{c^2} \frac{1}{\omega_{\mathrm{T}}^2 - \omega\left(\omega+ i \gamma\right) - \beta_{\mathrm{T}}^2 k^2} +  \frac{1}{\epsilon\left(\omega, k\right)}\right],
\end{align}
leading to the reflection coefficient
\begin{align}
	r = \frac{\frac{k_{\mathrm{B}}}{\epsilon_{\mathrm{B}}} - \frac{k_{\mathrm{T}}}{\epsilon\left(\omega,k\right)} - \Omega}{\frac{k_{\mathrm{B}}}{\epsilon_{\mathrm{B}}} + \frac{k_{\mathrm{T}}}{\epsilon\left(\omega,k\right)} + \Omega}.
\end{align}

\section{Additional Experimental Data}
\begin{figure}
	\includegraphics[width = \columnwidth]{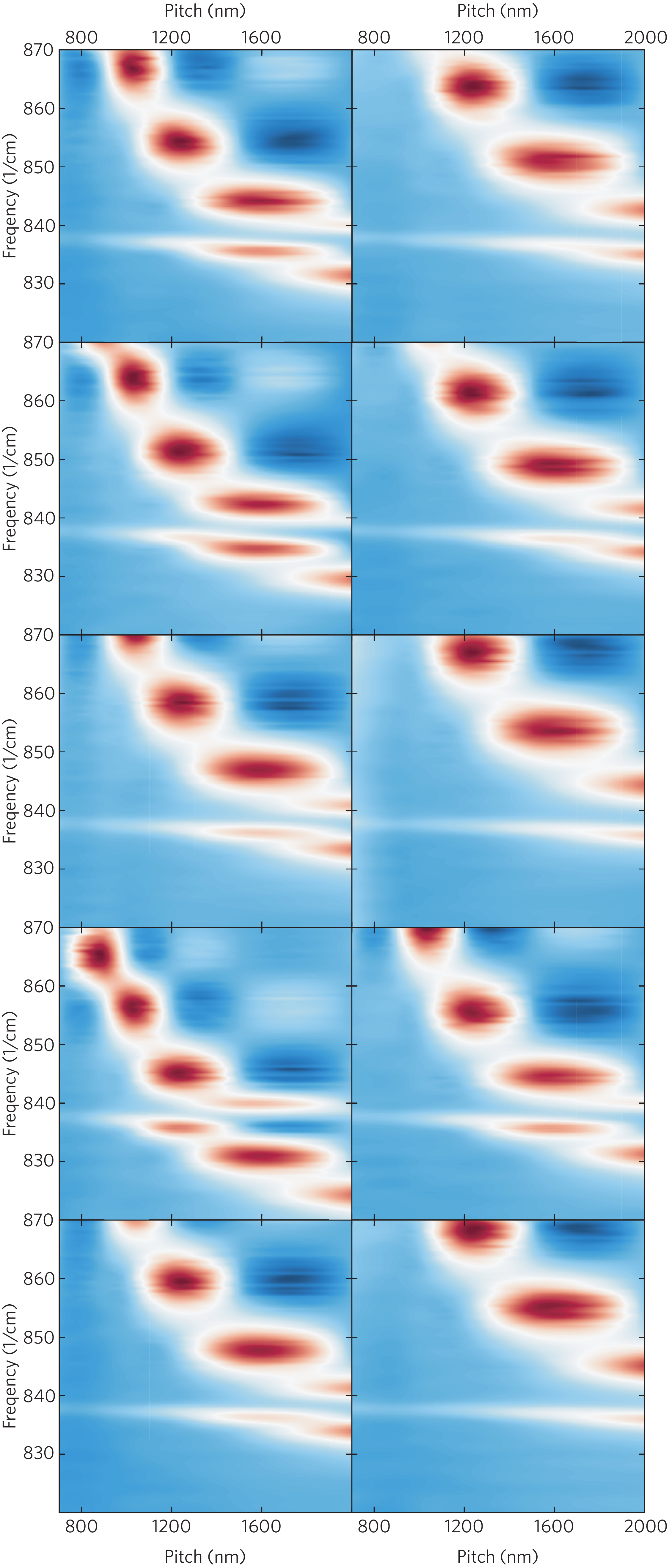}
	\caption{\label{fig:FigS1} Reflectance maps for samples of $300$nm (1st Column) and $500$nm (2nd Column). The rows correspond to different fabrication runs.}
\end{figure}

In the main body of the manuscript experimental data was presented from two samples with nominal pillar diameters of $300$nm and $500$nm. For each nominal pillar diameter 6 unique samples were fabricated. Due to variability in the fabrication procedure the frequencies of the optical modes varies slightly between samples. In Fig.~\ref{fig:FigS1} we show the reflectance spectra for the samples omitted from the main body of the text. Note that despite the variation in mode frequencies away from the weak LO phonon, every sample nonetheless reproduces the anticrossing.



\end{document}